\documentclass[doublecol]{epl2} 
\usepackage{amssymb}

\title{Transport coefficients of dissipative particle dynamics with finite 
time step}
\shorttitle{Transport coefficients of DPD with finite time step} 
\author{Hiroshi Noguchi\thanks{E-mail:\email{hi.noguchi@fz-juelich.de}} 
\and Gerhard Gompper}
\shortauthor{H. Noguchi \& G. Gompper}  %Insert here the first author if authors exceed 70 characters

\institute{
Institut f\"ur Festk\"orperforschung, Forschungszentrum J\"ulich, 
52425 J\"ulich, Germany
}


\abstract{
The viscosity and self-diffusion constant of a mesoscale hydrodynamic method, 
dissipative particle dynamics (DPD), are investigated.
The viscosity of DPD with finite time step, including the Lowe-Anderson 
thermostat, is derived analytically for the ideal-gas equation of state 
and phenomenologically for systems with soft repulsive potentials.
The results agree well with numerical data.  The scaling of the local 
relative velocity in molecular dynamics simulations is shown to be useful to 
obtain faster diffusion than for the DPD thermostat.
}

\begin{document}
\maketitle

\section{Introduction}
Soft matter systems such as polymer solutions, colloidal suspensions, vesicles, 
cells, and microemulsions exhibit many interesting dynamical 
behaviors, where hydrodynamic flow plays an important role, as do thermal 
fluctuations.
The characteristic length ($n$m to $\mu$m) and time ($n$s to s) scales of 
soft-matter systems are typically much larger than the atomistic scales.
Coarse-grained molecular models and simulation methods are therefore necessary
to simulate mesoscale phenomena  with reasonable computational effort.
Dissipative particle dynamics 
(DPD)~\cite{hoog92,groo97,shar03,pete04,alle06,boek97,spen00,jian07,vent06,mars97,espa99,mast99,ripo01,viss06}
has been developed for this purpose, and has been applied to various 
systems such as colloids \cite{boek97} 
polymers~\cite{groo97,spen00,jian07} and lipid membranes~\cite{vent06}.
DPD is an off-lattice hydrodynamic method,
which has two main features: soft-repulsive interaction potentials and
the pairwise version of a Langevin thermostat.
Since there is no impenetrable exclude volumes, 
a DPD particle describes not a solvent molecule but a fluid element, 
which represents clusters of solvent molecules.
The main motivation for the use of soft potentials is that they allows 
large time steps for the time evolution; however, it has been shown that
the simulations have to be checked carefully in each case 
by monitoring the configurational temperature~\cite{rugh97}, in 
order to avoid artifacts due to too large time steps~\cite{alle06}.
DPD shares many properties with direct simulation Monte Carlo~\cite{bird76} 
and multi-particle collision (MPC) dynamics 
\cite{male99,ihle01,kiku03,ripo04,hech05} as pointed out very 
recently in Ref.~\cite{nogu07}.

The transport coefficients of DPD have been studied 
for about a decade~\cite{hoog92,mars97,espa99,mast99,ripo01,viss06}.
However, the viscosity of DPD has been derived analytically only for the 
ideal-gas equation of state
in the small time-step limit~\cite{mars97}, and partially for  
the Anderson-thermostat~\cite{ande80} version proposed by Lowe~\cite{lowe99}. 
Recently, numerical integrators for the DPD thermostat~\cite{shar03,pete04}, 
which have no time-step dependence on thermodynamic properties,
were proposed; they include the Lowe-Anderson thermostat (Lowe-AT)
as a specific limit.
However, the transport coefficients do depend on the time step.
Therefore, a detailed understanding of the transport coefficients for
{\em finite time step} is very important to control and tune the hydrodynamic 
properties of DPD fluids.

In this letter, we calculate the time-step dependence of the viscosity, and 
in particular investigate the contribution due to the interaction potential.
The viscosity of DPD consists of three contributions, 
$\eta=\eta_{\rm {kin}}+\eta_{\rm {col}}+\eta_{\rm {pot}}$.
The kinetic viscosity $\eta_{\rm {kin}}$, collision viscosity 
$\eta_{\rm {col}}$, and potential viscosity $\eta_{\rm {pot}}$ 
result from the momentum transfer due to particle displacements, 
collisions generated the DPD thermostat (arising from frictional interactions 
and thermal noise), and potential interactions, respectively.
We determine these three contributions both analytically and numerically.
In previous DPD simulations with a repulsive potential,
the contributions of the potential interactions were often neglected
in the discussion of transport coefficients.
However, the potential contributes to the viscosity as well as the 
DPD thermostat in typical simulation conditions.
We also study the self-diffusion constant $D$ of a DPD particle and 
the ratio of momentum to mass transport,
which is characterized by the Schmidt number $Sc=\nu/D$, where 
$\nu=\eta/\rho$ is the kinematic viscosity.
Finally, we show that faster relaxation and
larger diffusion constants $D$ can be obtained in Molecular Dynamics (MD)
simulations
by the rescaling of the local relative velocity to control temperature
instead of a DPD thermostat.

\section{Methods}
The DPD thermostat is a modified Langevin thermostat,
where the friction and noise terms are applied to the relative velocities 
of the neighbor pairs. 
The equation of motion for the $i$-th particle with mass $m$ is given by
\begin{eqnarray} 
\label{eq:dpd}
m \frac{d {\bf v}_{i}}{dt} &=&
 - \frac{\partial U}{\partial {\bf r}_i} + f_{\rm {DT}}, \\ \nonumber
 f_{\rm {DT}} &=& \sum_{j\not=i} \left\{-w(r_{ij}){\bf v}_{ij}
   \cdot{\bf \hat{r}}_{ij} + 
      \sqrt{w(r_{ij})}{\xi}_{ij}(t)\right\}{\bf \hat{r}}_{ij},
\end{eqnarray} 
where ${\bf v}_{ij}={\bf v}_{i}-{\bf v}_{j}$, 
${\bf r}_{ij}= {\bf r}_{i}-{\bf r}_{j}$, ${r}_{ij}=|{\bf r}_{ij}|$, and
${\bf \hat{r}}_{ij}={\bf r}_{ij}/{r}_{ij}$.
The Gaussian white noise ${\bf \xi}_{ij}(t)$ 
obeys the fluctuation-dissipation theorem, with
 average $\langle \xi_{ij}(t) \rangle  = 0$ and variance
$\langle \xi_{ij}(t) \xi_{i'j'}(t')\rangle  =  
         2 k_{\rm B}T (\delta _{ii'}\delta _{jj'}+\delta _{ij'}\delta _{ij'}) 
\delta(t-t')$,
where $k_{\rm B}T$ is the thermal energy.
This thermostat is applied only in the direction ${\bf \hat{r}}_{ij}$
to conserve the local angular momentum.
In DPD, a linear weight function $w(r_{ij})=w_1(r_{ij})$, with 
\begin{equation}\label{eq:w1}
w_1(r_{ij})=\gamma\left(1-\frac{r_{ij}}{r_{\rm {cut}}}\right),
\end{equation}
is typically employed, which vanishes beyond the cutoff at 
$r_{ij}=r_{\rm {cut}}$.
Furthermore, DPD is usually combined with a soft repulsive 
potential~\cite{groo97}, 
\begin{equation}\label{eq:pot}
U=\frac{a k_{\rm B}T}{2} \sum_{i<j}
           \left(1-\frac{r_{ij}}{r_{\rm {cut}}}\right)^2, 
\end{equation}
with the same cutoff $r_{\rm {cut}}$, but other potentials are also 
available.

The DPD equation~(\ref{eq:dpd}) is discretized by
the Shardlow's S1 splitting algorithm~\cite{shar03}, where 
each thermostat of the $ij$ pair is separately integrated,
\begin{eqnarray}\label{eq:split}
{\bf v}_{i}^{\rm {new}} &=& {\bf v}_{i} + \{-A(r_{ij}){\bf v}_{ij} \nonumber
   \cdot{\bf \hat{r}}_{ij} + B(r_{ij}){\bf \xi}_{ij,n}\}{\bf \hat{r}}_{ij},\\
{\bf v}_{j}^{\rm {new}} &=& {\bf v}_{j} - \{-A(r_{ij}){\bf v}_{ij}
   \cdot{\bf \hat{r}}_{ij} + B(r_{ij}){\bf \xi}_{ij,n}\}{\bf \hat{r}}_{ij},
\end{eqnarray} 
with
\begin{equation}\label{eq:splitab}
A(r_{ij}) = \frac{w(r_{ij}) \Delta t/m}{1+w(r_{ij}) \Delta t/m},\  
B(r_{ij})= \frac{\sqrt{w(r_{ij})\Delta t}/m}{1+ w(r_{ij})\Delta t/m}.
\end{equation}
The discretized Gaussian noise ${\bf \xi}_{ij,n}$ is determined by 
$\langle \xi_{ij,n} \xi_{i'j',n'}\rangle  =  
2 k_{\rm B}T (\delta _{ii'}\delta _{jj'} 
                    + \delta _{ij'}\delta _{ij'}) \delta_{nn'}$.
This splitting algorithm belongs to the generalized Lowe-AT~\cite{pete04},
because the factors $A(r_{ij})$ and $B(r_{ij})$ satisfy the 
relation $B=\sqrt{A(1-A)/m}$~\cite{nogu07}.
Thus, for $U=0$ this algorithm yields the flat radial distribution function 
of an ideal gas for {\em any} time step $\Delta t$, without  
any deviation of the kinetic temperature from the thermostat temperature.
In the Lowe-AT~\cite{lowe99}, the relative velocity 
${\bf v}_{ij}\cdot{\bf \hat{r}}_{ij}$ 
of a neighbor pair $ij$ with $r_{ij}/r_{\rm {cut}}<1$ is updated
by assigning a random number drawn from the Maxwell-Boltzmann distribution 
with the probability $\Gamma'$ at each time step $\Delta t$
({\em i.e.} velocities are updated with the rate $\Gamma=\Gamma'/\Delta t$).
When a piecewise constant weight function $w(r_{ij})=w_0(r_{ij})$, where 
\begin{equation}
w_0(r_{ij})=\left\{
\begin{array}{ll}
\gamma  & {\rm for} \ \  r_{ij} < r_{\rm {cut}} \\ 
0       & {\rm otherwise},
\end{array}
\right.
\end{equation}
is employed, Eq.~(\ref{eq:split}) with $\gamma \Delta t/m=1$ gives the 
Lowe-AT for $\Gamma'=1$.

The viscosities are calculated from simulations of simple shear flow
in three dimensions with Lees-Edwards boundary conditions~\cite{alle87}.
We use the weight $w_1(r_{ij})$, defined in Eq.~(\ref{eq:w1}),
and the splitting algorithm (\ref{eq:split}) for the DPD simulations. 
However, the derived analytical expressions can be applied to other 
weights $w(r_{ij})$ and other generalized Lowe-AT algorithms such as 
$A(r_{ij})$ in table I of Ref.~\cite{pete04}.
The self-diffusion constant $D$ is calculated from the mean square 
displacement of a particle,
$\langle \{{\bf r}_i(t)-{\bf r}_i(0)\}^2\rangle = 2d D t$,
where $d$ is the spatial dimension.

We have performed simulations with the usual soft potential (\ref{eq:pot}) 
in order to investigate the effect of interaction potentials.
The multi-time-step algorithm~\cite{pete04,tuck92} is employed, with a 
shorter time step $\delta t$ for the force $-\partial U/\partial {\bf r}_{i}$, 
so that the configurational\cite{alle06,rugh97} and thermostat temperatures 
 differ by less than $0.5$\%.
The side lengths of the simulation box are $L_y\geq 40r_{\rm {cut}}$, 
$L_x=L_z=10r_{\rm {cut}}$ and $L_x=L_y=L_z=20r_{\rm {cut}}$ for the 
calculation  of the viscosity and the diffusion constant, respectively.
The error bars of the simulation results are estimated from three 
independent runs.
We display our simulation results in form of dimensionless quantities,
indicated by a superscript, $\gamma^*=\gamma\tau_{\rm 0}/m$,
$\Delta t^*=\Delta t/\tau_{\rm 0}$,
$\delta t^*=\delta t/\tau_{\rm 0}$,
and the number density $n^*=n {r_{\rm {cut}}}^d$, which corresponds to 
measuring 
length, time, viscosity, and diffusion constant of a particle in units
of $r_{\rm {cut}}$, $\tau_{\rm 0}=r_{\rm {cut}}\sqrt{m/k_{\rm B}T}$,
$\eta_{\rm 0}=\sqrt{mk_{\rm B}T}/{r_{\rm {cut}}}^{d-1}$, and
$D_{\rm 0}= r_{\rm {cut}}\sqrt{k_{\rm B}T/m}$, respectively.

\section{Viscosity of an ideal DPD gas}

First, we derive the viscosity of an ideal gas of DPD particles, with 
$U = 0$ and $\eta_{\rm {pot}}=0$, using a kinetic-theory approach.
In simple shear flow with flow velocity ${\bf v}=\dot\gamma y {\bf e}_x$,
the $xy$ component of the stress tensor is given by 
$\sigma_{xy}=\eta \dot\gamma$.
The viscosities $\eta_{\rm {kin}}$ and $\eta_{\rm {col}}$ are
calculated from the stress due to the kinetic and collisional contributions, 
respectively.

\begin{figure}
\onefigure{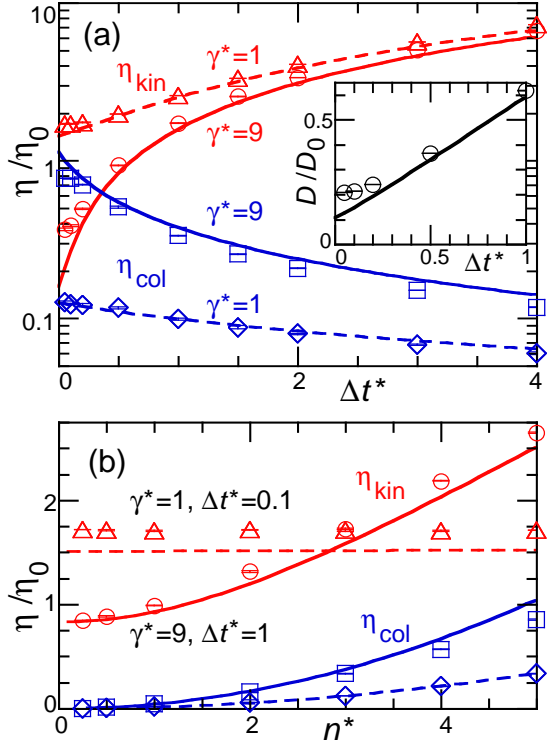}
\caption{
(Color online)
Dependence of the viscosity $\eta$ of an ideal DPD gas 
(with $U=0$) on (a) the time step $\Delta t^*$ at $n^*=3$
and (b) the number density $n$ for $\Delta t^*=1$.
Symbols indicate simulation data for $\gamma^*=1$ ($\triangle$, $\diamond$)
and $\gamma^*=9$ ($\circ$, $\Box$). 
Lines represent the analytical results of Eqs.~(\ref{eq:kinv0}) and 
(\ref{eq:colv0}).  The inset in (a) shows the dependence of the diffusion 
constant $D$ of the ideal DPD gas on
the time step $\Delta t^*$ for $n^*=3$ and $\gamma^*=9$.}
\label{fig:ig}
\end{figure}

The kinetic stress $\sigma_{xy}^{\rm {kin}}$ is the momentum flux due to 
particles crossing a plane of constant $y$; it 
can be calculated by following the derivation for MPC in Ref.~\cite{kiku03}.
The stress is written as 
\begin{eqnarray}\label{eq:stk}
\sigma_{xy}^{\rm {kin}} &=& \frac{mn}{\Delta t} \biggl\{-\int_{-\infty}^{0} dy 
        \int_{v_y>- \frac{y}{\Delta t}} d{\bf v}\ v_x 
                P({\bf v}-\dot\gamma y {\bf e}_x) \nonumber \\
&&+ \int_{0}^{\infty} dy \int_{v_y<-\frac{y}{\Delta t}} d{\bf v}\ v_x 
                P({\bf v}-\dot\gamma y {\bf e}_x) \biggr\},
\end{eqnarray}
where $P({\bf v})$ is the velocity probability distribution in the local 
rest frame.
This stress can be rewritten as $\sigma_{xy}^{\rm {kin}}=
mn(\dot\gamma \Delta t \langle v_x^2 \rangle/2 - \langle v_x v_y \rangle)$.
The velocity distribution is shifted by particle streaming in the time
interval $\Delta t$, so that
$\langle v_x v_y \rangle \rightarrow 
\langle v_x v_y \rangle - \dot\gamma\Delta t\langle v_x v_y \rangle$.
Then, the DPD collisions of Eq.~(\ref{eq:split}) modify it as $\langle v_x v_y \rangle \rightarrow  s \langle v_x v_y \rangle$.
Thus, the self-consistency condition of a stationary shear flow is 
$\langle v_x v_y \rangle= s (\langle v_x v_y \rangle - 
\dot\gamma\Delta t\langle v_x v_y \rangle)$.
The kinetic viscosity $\eta_{\rm {kin}}$ is then given by~\cite{kiku03}
\begin{equation}\label{eq:kinv0}
\eta_{\rm {kin}}= nk_{\rm B}T\Delta t\left( \frac{1}{1-s}-\frac{1}{2} \right).
\end{equation}
The remaining task is to calculate the factor $s$ for the DPD collisions.
The $i$-th particle collides with a multitude of other particles at the 
same time step, so that $s=\langle\Pi_j s_{ij}\rangle$.
Eq.~(\ref{eq:split}) together with a molecular chaos assumption implies 
$s_{ij}= 1- A({\hat{x}_{ij}}^2+{\hat{y}_{ij}}^2) 
      + 4A^2{\hat{x}_{ij}}^2{\hat{y}_{ij}}^2$,
where $\hat{x}_{ij}$ and $\hat{y}_{ij}$ are the components of 
${\bf \hat{r}}_{ij}$. In an ideal gas,
the local number density fluctuates around the average $n$, and
the number of particles $k$ per volume $\Delta V$ is given by the Poisson 
distribution, $P(k)= e^{-n\Delta V}(n\Delta V)^k/k!$, which implies 
$\langle c^k\rangle=\exp\{(-1+c)n\Delta V\}$ for some constant $c$.
Therefore, the factor $s$ is given by
\begin{eqnarray}\label{eq:f}
s &=& \exp\left\{n \int \left(- 2A(r)\hat{x}^2 + 4A(r)^2\hat{x}^2\hat{y}^2
                            \right) dV\right\} \\ \nonumber
  &=& \exp\left\{n \int \left(-\frac{2A(r)}{d}+ \frac{4A(r)^2}{d(d+2)} 
                            \right) dV\right\}.
\end{eqnarray}
Eqs.~(\ref{eq:kinv0}) and (\ref{eq:f}) give the kinetic viscosity 
$\eta_{\rm {kin}}$ for a finite time step $\Delta t$.
In the continuum limit $\Delta t \ll 1$, 
we recover the result 
\begin{equation}\label{eq:kinv1}
\eta_{\rm {kin}}= \frac{d m k_{\rm B}T}{2[w]_g},
\ \ [w]_g \equiv \int g(r)w(r) dV
\end{equation}
of Ref.~\cite{mars97},
where $g(r)$ is the radial distribution function, with $g(r)=1$ for the 
ideal gas.
In the Lowe-AT, the factor $s$ is given by $s=\exp(-\pi b n\Gamma')$
with $b=1/2$ and $b=16/45$ in two and three spatial dimensions, respectively.
In the limit $\Delta t \ll 1$ with finite $\Gamma$, 
$\eta_{\rm {kin}}=k_{\rm B}T/\pi b \Gamma$.

\begin{figure}
\onefigure{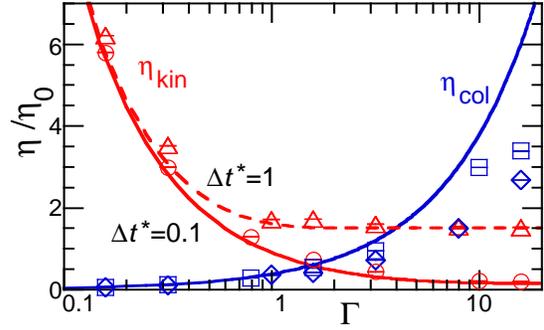}
\caption{
(Color online)
Dependence of the viscosity of an ideal DPD gas 
with Lowe-Anderson thermostat on
the normalized collision frequency $\Gamma$.
Symbols represent simulation data at $\Delta t^*=0.1$ ($\circ$, $\Box$)
and $\Delta t^*=1$ ($\triangle$, $\diamond$).
Lines indicate the analytical results.}
\label{fig:la}
\end{figure}

The collisional stress $\sigma_{xy}^{\rm {col}}$ is the momentum flux due 
to DPD collisions --- determined by Eq.~(\ref{eq:split}) --- crossing a 
plane at $y=y_0=0$,
\begin{equation}\label{eq:colst}
\sigma_{xy}^{\rm {col}}= -n^2 \int_{0}^{\infty} dy_i \int_{y_{ij}>y_i}  
d{\bf r}_{ij} \frac{m(v_{i,x}^{\rm {new}}-v_{i,x})}{\Delta t}.
\end{equation}
After substitution of Eq.~(\ref{eq:split}) and  
$\langle  v_{ij,x}\rangle=\dot\gamma y_{ij}$ into Eq.~(\ref{eq:colst}) and 
interchange of the order of integration,
$\eta_{\rm {col}}$ is found to be 
\begin{eqnarray}\label{eq:colv0}
\eta_{\rm {col}} &=& \frac{n^2}{2} \int d{\bf r} 
       \frac{A(r)m r^2 \hat{x}^2\hat{y}^2}{\Delta t}  \nonumber \\
 &=& \frac{n^2}{2d(d+2)} \left[\frac{w r^2}{1+w\Delta t/m} \right]_g.
\end{eqnarray}
Equation (\ref{eq:colv0}) gives $\eta_{\rm {col}}=\{n^2/2d(d+2)\}[w r^2]_g$ 
in the limit $\Delta t \ll 1$. For the Lowe-AT, Eq.~(\ref{eq:colv0})
implies $\eta_{\rm {col}}=\pi m n^2\Gamma {r_{\rm {cut}}}^4/64$ and 
$\eta_{\rm {col}}=\pi m n^2\Gamma {r_{\rm {cut}}}^5/75$  
in two and three spatial dimensions, respectively.
These results agree with the collisional viscosities obtained in 
Refs.~\cite{mars97} and \cite{lowe99}.

The analytical results agree well with the numerical data, 
as shown in Figs.~\ref{fig:ig} and \ref{fig:la}.
As $\Delta t$ increases, $\eta_{\rm {kin}}$ increases but $\eta_{\rm {col}}$ 
decreases, just like the viscosities of MPC-Langevin dynamics~\cite{nogu07}.
Although $\eta_{\rm {kin}}$ is almost independent of the density $n$ at 
small $\Delta t$, $\eta_{\rm {kin}}$ increases with $n$ at large $\Delta t$, 
see Fig.~\ref{fig:ig}(b).
There are small deviations between analytical and numerical results in 
Fig.~\ref{fig:ig}.  They are of the same order of magnitude as the 
deviations for $\Delta t \ll 1$ reported in Ref.~\cite{mast99}, which
have been explained by correlation effects between DPD collisions~\cite{mast99}.
In the Lowe-AT, the viscosities depend on $\Delta t$ for large $\Gamma$,
see Fig.~\ref{fig:la}.
For $\Gamma'=\Gamma\Delta t > 1$, our theory overestimates $\eta_{\rm {col}}$,
since the relative velocities of some $ij$ pairs are updated more than 
once in one time step.

\begin{figure}
\onefigure{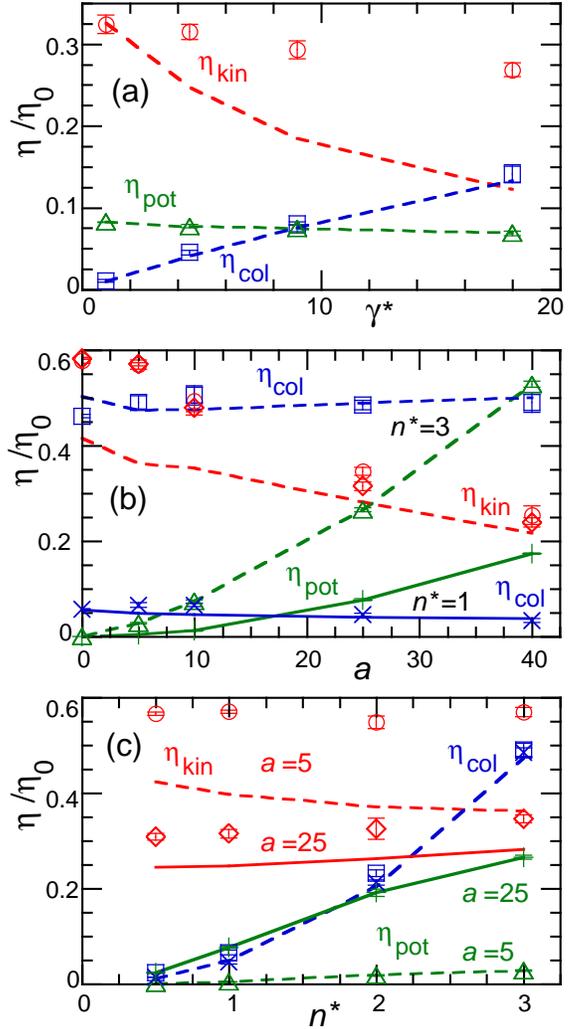}
\caption{
(Color online)
Dependence of the viscosity of DPD with soft potentials, Eq.~(\ref{eq:pot}), on
(a) friction coefficient $\gamma^*$ for $n^*=1$ and $a=25$,  
(b) potential strength $a$ for $\gamma^*=4.5$, and (c) number density $n$ for 
$\gamma^*=4.5$. In all cases,  $\delta t^*=0.01$, and $\Delta t^*=0.1$.
Symbols indicate simulation data for 
(b) $n^*=3$ ($\circ$, $\Box$, $\triangle$) and $n^*=1$ ($\diamond$, $\times$, $+$), and 
(c) $a=5$ ($\circ$, $\Box$, $\triangle$) and $a=25$ ($\diamond$, $\times$, $+$).
Lines for $\eta_{\rm {col}}$ represent the analytical results of 
Eq.~(\ref{eq:colv0}).
Lines for $\eta_{\rm {kin}}$ show the results of Eq.~(\ref{eq:kinv2})
with $\phi_{\rm {pot}}$ fitted by Eq.~(\ref{eq:potv1}).
Lines for $\eta_{\rm {pot}}$ are guides to the eye.}
\label{fig:dp}
\end{figure}

\section{Viscosity with interaction potential}

With interaction potential,
an additional momentum flux crossing a plane at $y=0$,
is caused by the forces $f(r_{ij})= -\partial U/\partial r_{ij}$ 
between $ij$ pairs with $y_i>0$ and $y_j<0$.
The potential viscosity $\eta_{\rm {pot}}$ is given by
\begin{eqnarray}\label{eq:potv0}
\eta_{\rm {pot}} &=& -\frac{n^2}{\dot\gamma} 
    \int_{0}^{\infty} dy_i \int_{y_{ij}>y_i} 
          d{\bf r}_{ij}\ g(r_{ij})f(r_{ij})\hat{x}_{ij} \nonumber \\
&=&  - \frac{n^2}{2\dot\gamma} \int dV\ g(r)f(r)\hat{x}y,
\end{eqnarray}
which is the potential term of the Irving-Kirkwood formula of the 
viscosity~\cite{irvi50}.  The potential also modifies 
$\eta_{\rm {kin}}$ with an additional velocity relaxation, while 
$\eta_{\rm {col}}$ can be calculated by Eq.~(\ref{eq:colv0}) with 
non-uniform $g(r)$.  The viscosity with an interaction potential
has been derived analytically for some cases~\cite{resi87},
but is generally very complicated. Therefore, 
we employ a simple phenomenological expression instead, and focus on
the explanation of qualitative dependences.

In the molecular-chaos approximation, the velocity auto-correlation function 
of a particle in a gas shows
an exponential decay, $\langle v(t)v(0)\rangle=\exp(-\phi t)$.
This behavior corresponds to the assumption of a Langevin equation;
$dv/dt=-\phi v + \sqrt{\phi}\xi(t)/m$
for the particle velocity $v$ in the local rest frame.
For $\Delta t \ll 1$, the DPD collisions generate an  
auto-correlation function with an initial exponential decay (for small 
times $t$) with $\phi_{\rm {DPD}}=n[w]_g/dm$.
Here, we assume that the potential also generates an exponential 
auto-correlation function with rate $\phi_{\rm {pot}}$,
although the auto-correlation function determined numerically is {\it not}
exponential, and shows larger deviation from an exponential decay for 
larger potential strengths $a$ or particle densities $n$.
Then, the kinetic viscosity $\eta_{\rm {kin}}$ is given by
\begin{equation}\label{eq:kinv2}
\eta_{\rm {kin}}= \frac{n k_{\rm B}T}{2(\phi_{\rm {pot}} + \phi_{\rm {DPD}})},
\end{equation}
compare Eq.~(\ref{eq:kinv1}).

In order to estimate $\eta_{\rm {pot}}$, an expression for the correlations 
of $ij$ pairs is required.
We mimic the potential contribution by a DPD thermostat,
\begin{equation}\label{eq:potdpd}
m\frac{d {\bf v}_{i}}{dt} =
 \sum_{j\not=i} \left\{-\gamma_{\rm {pot}}|f(r_{ij})|{\bf v}_{ij}
   \cdot{\bf \hat{r}}_{ij} + 
      \sigma{\xi}_{ij}(t)\right\}{\bf \hat{r}}_{ij},
\end{equation}
where $\sigma=\sqrt{\gamma_{\rm {pot}}|f(r_{ij})|}$,
since the restoring force should be proportional to $|f(r_{ij})|$, and 
$v_{ij}$ in the direction ${\bf \hat{r}}_{ij}$. 
Following the derivation of Eq.~(\ref{eq:colv0})
with $\phi_{\rm {pot}}=n\gamma_{\rm {pot}}[|f|]_g/dm$,
we obtain the viscosity 
\begin{equation}\label{eq:potv1}
\eta_{\rm {pot}} = \frac{n m \phi_{\rm {pot}} [|f| r^2]_g}{2(d+2)[|f|]_g}.
\end{equation}
Thus, the viscosities $\eta_{\rm {kin}}$ and $\eta_{\rm {pot}}$ can be 
estimated by Eqs.~(\ref{eq:kinv2}) and (\ref{eq:potv1}) with the parameter 
$\phi_{\rm {pot}}$ and the radial distribution function $g(r)$.

Figure~\ref{fig:dp} shows the viscosities of the DPD fluid with interaction potential.
 We calculate $g(r)$ from equilibrium simulations, 
fit $\phi_{\rm {pot}}$ to $\eta_{\rm {pot}}$,
and then estimate $\eta_{\rm {kin}}$ from Eq.~(\ref{eq:kinv2}).
This underestimates $\eta_{\rm {kin}}$, but reproduces very well the 
qualitative dependence on friction coefficient $\gamma$, potential strength
$a$, and number density $n$.  
The kinetic viscosity $\eta_{\rm {kin}}$ decreases with 
increasing $\gamma$ or $a$. The potential viscosity $\eta_{\rm {pot}}$ is 
almost independent of $\gamma$ and increases with $a$.
The collision viscosity $\eta_{\rm {col}}$ is almost independent of
$a$ and shows very good agreement between the theory and simulations.

\section{Diffusion}

Next, we derive the self-diffusion constant $D$ of an ideal gas of DPD 
particles (with $U=0$) for finite time steps.
Following the derivation of Eq.~(\ref{eq:f}),
we find that the velocity correlation for one step is given by
$\langle v_x(t+\Delta t) v_x(t)\rangle = \exp(-n[A]_g/d)$.
Under the molecular chaos assumption, i.e. 
$\langle v_x(k \Delta t) v_x(0)\rangle = \langle v_x(\Delta t) v_x(0)\rangle^k$,
the diffusion constant is then given by
\begin{equation}\label{eq:dif0}
D= \frac{k_{\rm B}T\Delta t}{m} 
           \left( \frac{1}{1-\exp(-n[A]_g/d)} - \frac{1}{2} \right).
\end{equation}
In the limit $\Delta t \ll 1$, the diffusion constant becomes 
$D= d k_{\rm B}T/n [w]_g$, in agreement with the result of Ref.~\cite{mars97}.
However, the velocity auto-correlation function 
$\langle v_x(k \Delta t) v_x(0)\rangle$ with large dimensionless friction 
coefficient $\gamma^*$ has a long-time tail due to the hydrodynamic 
interactions~\cite{espa99}, and the diffusion constant $D$ becomes larger 
than the value in Eq.~(\ref{eq:dif0}).
This underestimation of $D$ is seen at small $\Delta t$
in the inset of Fig.~\ref{fig:ig}(a).

Since the kinetic contribution to the kinematic viscosity 
$\eta_{\rm {kin}}/\rho$ is roughly proportional to $D$,
the relation $\eta_{\rm {kin}} \ll \eta_{\rm {col}}+\eta_{\rm {pot}}$ at 
large friction coefficient $\gamma^*$ or potential strength $a$ yields 
large Schmidt numbers $Sc$ in DPD.
On the other hand, small $\gamma^*$ and $a$ gives $Sc<1$, {\it e.g.}
$Sc=\eta_{\rm {kin}}/\rho D=1/2$ for $\Delta t \ll 1$ and 
$\eta_{\rm {kin}} \gg \eta_{\rm {col}}+\eta_{\rm {pot}}$.
Sufficiently large $Sc$ yields hydrodynamic behavior.
For example, a large Schmidt number is required in polymer simulations
to produce Zimm dynamics \cite{doi86} --- where the relaxation time $\tau_p$ 
of a mode with mode number $p$ is expected to scale as 
$\tau_p\sim (N_m/p)^{3/2}$ --- with moderate chain lengths $N_m$
for an ideal chain, as 
demonstrated in MPC simulations with $Sc \simeq 10$~\cite{ripo04}.
Zimm dynamics was also reported from DPD simulations with the most 
typical parameters $n^*=3$, $a=25$ and $\gamma^*=4.5$~\cite{jian07} or 
$5.6$~\cite{spen00}; however, the variation of the Zimm exponent  
with temperature observed in Ref.~\cite{jian07}
seems to indicate that the simulations were performed in the region  
between the Rouse and Zimm regimes.  From our results above, we obtain
$Sc=1.7$ and $\eta_{\rm {col}}+\eta_{\rm {pot}} \simeq 2\eta_{\rm {kin}}$
at $n^*=3$, $a=25$, $\gamma^*=4.5$ with $\delta t=0.01$ and $\Delta t=0.1$.
Thus, this parameter set is indeed in the crossover region between 
gas-like and liquid-like behavior.

\section{Other Thermostats}
To simulate the hydrodynamic behavior of complex fluids,
dimensionless hydrodynamic quantities, such as
the Reynolds number and the Schmidt number $Sc$,
typically have to be adjusted to match experimental conditions.
To study low-Reynolds-number flows of soft matter and biological systems, 
high viscosity is often required.
On the other hand, DPD simulations are also often employed to study 
equilibrium properties. In this case, faster diffusion and 
lower viscosity is advantageous, since it provides faster relaxation into 
the equilibrium state.
Recently, a Nos{\'e}-Hoover-type thermostat for the relative velocities 
of neighbor pairs was proposed~\cite{groo05,alle06a},
where the momentum is locally conserved.
Its main idea is to thermostat systems,
but to less disturb the original hydrodynamic transport properties (in the
absence of any thermostat).
However, the Nos{\'e}-Hoover thermostat usually has to be combined with 
another thermostat to keep the temperature constant,
when a system includes a potential with strong $C^2$ discontinuity like 
Eq.~(\ref{eq:pot}).

\begin{figure}
\onefigure{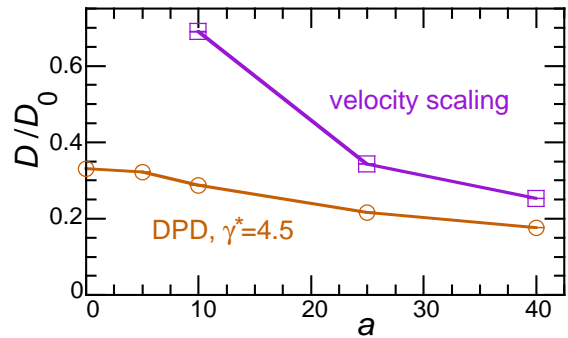}
\caption{
(Color online)
Dependence of the diffusion constant $D$ on the strength $a$ of the 
soft potential (\ref{eq:pot}), 
for $n^*=3$, $\delta t^*=0.01$, and $\Delta t^*=0.1$.
The velocity rescaling is performed with time step $\Delta t^*$ and cell size $l_{\rm c}=r_{\rm {cut}}$.
In comparison, the diffusion constant $D$ of DPD with $\gamma^*=4.5$ 
is also shown.}
\label{fig:vs}
\end{figure}

The scaling of velocities~\cite{alle87} is an easy way to control the 
temperature in MD simulations in thermal equilibrium. 
In order to retain hydrodynamic properties, {\it e.g.} under flow, 
the main issue is  momentum conservation,
{\it i.e.}  how to determine the velocity of the local rest frame.
We suggest to employ the velocity scaling of the MPC method~\cite{male99},
which can be used independent of the MPC collision procedure.
The particles are sorted into the cells of a cubic lattice with lattice 
constant $l_{\rm c}$, and the local flow velocity is identified with the 
velocity ${\bf v}_{\rm c}^{\rm G}$ of the center of mass of all particles 
in a cell.  Then, the relative velocities 
${\bf u}_{i}={\bf v}_{i}-{\bf v}_{\rm {c}}^{\rm G}$ are rescaled as
${\bf u}_{i} \rightarrow {\bf u}_{i}\sqrt{d(N-N_{\rm {c}})k_{\rm B}T/m\sum_i {{\bf u}_{i}}^2}$,
where $N$ is the total number of particles and 
$N_{\rm {c}}$ is the number of cells occupied by particles.
The cells are randomly shifted before each scaling step to ensure  
Galilean invariance~\cite{ihle01}.
The velocity scaling gives faster diffusion than DPD as shown in 
Fig.~\ref{fig:vs}. This is particularly important for solvents with 
Lennard-Jones-type interactions, where the frictional contributions of 
a DPD thermostat adds up with an already high viscosity in classical MD.
Velocity rescaling can produce temperature gradients in flow due to a
locally inhomogeneous energy dissipation.
To reduce these gradients,
a local version of rescaling procedure can be employed.
To do so, many cells are grouped into larger bins, typically arranged
sequentially in layers or columns,
and the rescaling is performed individually for each bin.
Alternatively, gradients on the cell scale can be avoided
by velocity scaling with a Monte Carlo scheme~\cite{hech05},
where the scaling factor fluctuates stochastically in each cell to reproduce
the correct kinetic energy distributions.

\section{Summary}

We have studied the viscosity of DPD with finite time step, both analytically 
and numerically. The analytical results agree very well with the simulation 
data.  Our theoretical results for the viscosity can be generalized 
straightforwardly to other DPD methods, such as DPD with a multibody 
thermostat~\cite{nogu07}. Thus, we have shown that by varying the time step
$\Delta t$ and the friction coefficient $\gamma$, the dynamic properties 
of a DPD solvent can be tuned, while thermodynamic properties remain 
unaffected.

Furthermore, we have shown that the velocity 
rescaling method, which is routinely employed in MPC, can be adapted to
MD simulations. It respects Galilean invariance and disturbs the original 
hydrodynamics much less than a DPD thermostat.

\acknowledgments
We thank M.~Ripoll for helpful discussions. 
The partial support of this work by the DFG 
through the priority program ``Nano- and Microfluidics'' is gratefully
acknowledged.

\end{document}